\documentclass[]{JHEP3}
\usepackage{epsf,graphicx}
\title{Brane-World Cosmology, Bulk Scalars and Perturbations}
\author{Philippe Brax\footnote{On leave of absence from
Service de Physique Th\'eorique, CEA-Saclay, F-91191 Gif/Yvette cedex France}\\Theoretical Physics 
Division, CERN\\ CH-1211 Geneva 23.\\
{\tt philippe.brax@cern.ch}}
\author{Carsten van de Bruck \& Anne C. Davis\\ 
DAMTP, Centre for Mathematical Sciences\\ 
Cambridge University, Wilberforce Road, Cambridge,
CB3 0WA, UK.\\
{\tt C.VanDeBruck@damtp.cam.ac.uk, a.c.davis@damtp.cam.ac.uk}}
\abstract{We investigate aspects of cosmology in brane world theories 
with a bulk scalar field.
We concentrate on a recent model motivated from supergravity in
singular spaces. After discussing the background evolution of such a
brane-world, we present the evolution of the density contrast. We
compare our results to those obtained in the (second) 
Randall--Sundrum scenario and usual 4D scalar--tensor theories.}
\preprint{CERN-TH/2001-225\\DAMTP-2001-51}
\begin{document}
\section{Introduction}
The idea that we live on a hypersurface embedded in a higher 
dimensional space has sparked a lot of interest recently. 
These theories are motivated from string theory, 
where higher--dimensional objects, such as D--branes, play an 
essential role \cite{Polchinski}. Similarily, compactifying M--theory (or  
its effective low--energy limit: 11--D supergravity) on a $S_1/Z_2$--orbifold
and compactifying six dimensions on a Calabi--Yau manifold, 
results in a five--dimensional brane world scenario with 
two hypersurfaces, each located at the orbifold fixed points
(see e.g. \cite{HW1} and \cite{HW2}). 

Brane world theories predict that our universe was
higher--dimensional in the past. Because of this, there is the hope that 
certain questions which cannot be answered within the context 
of the standard model of cosmology, can be addressed within 
these theories. Furthermore, cosmology should be a way to severely 
constrain parameters in these models. 

So far, most cosmological considerations of brane worlds centered 
around the one--brane scenario of Randall and Sundrum \cite{RSII}. 
In this model, the three--dimensional brane universe is embedded in a 
5--dimensional Anti--de Sitter (AdS) spacetime. In particular, 
the bulk--space is empty, the only contribution to the curvature 
comes from the negative cosmological constant in the bulk. This 
simple model already leads to new effects which are interesting 
for cosmology \cite{RScosmo}. However, 
most scenarios motivated from particle physics predict matter in the 
bulk, such as scalar fields. In 5D heterotic M--theory, for
example, one particular scalar field measures the deformation 
of the Calabi--Yau manifold, on which six other small dimensions 
are compactified \cite{HW2}. Other models, motivated by supergravity (SUGRA), 
also predict bulk matter, whose form is dictated by the field theory under
consideration. 

Cosmology may be a fruitful field where the above ideas
can be tested. As such, the study of the evolution of
cosmological perturbations is extremely important \cite{robert}, 
because the higher--dimensional nature of the world can leave  traces in 
the distribution of matter and/or anisotropies in the microwave 
sky. A lot of papers investigated different aspects of perturbations 
in brane world scenarios \cite{braneworldperturbations}.

The aim of this paper is to develop an understanding of the evolution 
of perturbations in brane world scenarios, in which scalar field(s)
are present in the bulk. As a toy model, we will use a cosmological 
realization of a supergravity model in singular spaces
\cite{singular}. The evolution
of the brane world was discussed in depth in \cite{BDI} and 
\cite{BDII} (see also \cite{bulkscalar} for a discussion on 
brane cosmology and bulk scalar fields). In particular four 
different cosmological eras have been identified in this model. 
At high energy above the brane tension the cosmology is
non-conventional before entering the radiation epoch where the scalar
field
is frozen. After matter-radiation equality the scalar field starts
evolving in time leading to a slow-down of the expansion rate compared
to FRW cosmology in the matter dominated era. Eventually the scalar
field dynamics becomes the dominant one leading to a supergravity era.
Requiring that coincidence between the matter and
scalar field
energy densities  occurs in the recent past leads a cosmological
constant with a fine-tuning of the supersymmetry breaking
tension on the brane. 
In the supergravity era it has been  shown that the observed
acceleration of the expansion of the universe can be understood, i.e.
the computed acceleration paremeter $q_0=-4/7$ is within the
experimental ball-park. 
This accelerated expansion is driven by a bulk scalar 
field, whose parameters are constrained by the gauged supergravity  theory 
in the bulk. On top of this, the model  predicts a significant evolution
of the induced gravitational constant $G$ on the brane. Indeed,  
 the value of $G$ can be seen to have  changed by 37 percent since
radiation/matter equality. Thus, the study of cosmological
perturbation theory in these models is rather important, in particular
the time evolution of $G$ may leave  traces in the evolution of 
perturbations. In this paper we  discuss the evolution of
the density constrast and the effect of the bulk scalar field on 
cosmological perturbations.  In Section
2 we review Einstein's equations induced on our brane world 
and discuss the Friedmann equation. We also review the background
solutions found in \cite{BDII}, which are needed in order 
to solve the perturbation equations. In Section 3 we derive the
perturbation equations using the fluid flow approach. This approach is 
very transparent for our purposes and makes it easy to derive the 
necessary evolution equations. In Section 4 we discuss some solutions of these 
equations and discuss their properties. We point out the differences
to the Randall--Sundrum scenario and usual four--dimensional scalar--tensor 
theories. We conclude in Section 5. In the appendix we discuss some 
details concerning the issue of supersymmetry breaking and conformal 
flatness.

\section{Brane Cosmology}
In this section we discuss the field equations on the brane and
discuss the background evolution. 

\subsection{The Background Evolution}

 We consider our universe to be a boundary of a five dimensional space-time.
The embedding is chosen such that our brane-world sits at the origin of the fifth dimension.
We impose a $Z_2$ symmetry along the fifth dimension and identify $x_5$ with $-x_5$. 
Our brane-world carries two types of matter, the standard model fields at sufficiently large energy
and ordinary matter and radiation at lower energy. We also assume that gravity propagates
in the bulk where a scalar field $\phi$ lives. This scalar field couples to the standard model fields
living on the brane-world. At low energy when the standard model fields have condensed and
the electro-weak and hadronic phase transitions have taken place, the coupling of the scalar field
to the brane-world realizes the mechanism proposed in \cite{self-tune} with a self-tuning
of the brane tension.  
In this section we derive the brane cosmology equations describing the
coupling between ordinary matter on the brane and a scalar field in the bulk.

Consider the bulk action
\begin{equation}
S_{bulk}=\frac{1}{2\kappa_5^2}\int d^5 x \sqrt {-g_5}(R-\frac{3}{4}((\partial\phi)^2+U(\phi)))
\end{equation}
where $\kappa_5^2=1/M_5^3$ and the boundary action
\begin{equation}
S_{B}=-\frac{3}{2\kappa_5^2}\int d^4 x \sqrt{ -g_4}U_B(\phi_0)
\end{equation}
where $\phi_0$ is the boundary value of the scalar field.
The Einstein equations read 
\begin{equation}
G_{ab}\equiv R_{ab}-\frac{1}{2}R g_{ab}=T_{ab} +\delta_{x_5}T^B_{ab}
\end{equation}
where $T_{ab}$ is the bulk energy-momentum tensor and $T^B_{ab}$
is the boundary contribution. 
The bulk term is
\begin{equation}
T_{ab}=\frac{3}{4}\left(\partial_{a}\phi\partial_{b}\phi -\frac{1}{2}g_{ab}(\partial\phi)^2\right)
-\frac{3}{8}g_{ab}U
\end{equation}
and the boundary term
\begin{equation}
T^B_{ab}=-\frac{3}{2} g_{ab}U_B(\phi)
\end{equation}
with $a,b=0\dots 3$ in the last equation.
Following the self-tuning proposal we interpret $U_B$ as arising from a direct coupling
$U_B^0$ to the brane degrees of freedom, i.e. the standard model fields $\Phi^i$. The vacuum energy generated
by the $\Phi^i$'s yields the effective coupling
\begin{equation}
\frac{3U_B}{2\kappa_5^2}=<V(\Phi)>U_B^0
\end{equation}
where the dimension four potential  $V(\Phi)$ represents all the contributions due to the fields $\Phi^i$ after
inclusion of condensations, phase transitions and radiative corrections. 

We also consider that ordinary matter lives on the brane with a diagonal energy momentum tensor
\begin{equation}
\tau^{a}_{~b}=\hbox{diag}(-\rho,p,p,p)
\end{equation}
and an equation of state $p=\omega \rho$.

We will be mainly concerned with models derived from supergravity in
singular spaces. They involve $N=2$ supergravity with vector multiplets.
When supergravity in the bulk couples to the boundary in a supersymmetric way 
the Lagrangian is entirely specified by the superpotential
\begin{equation}
U_B=W
\end{equation}
and the bulk potential
\begin{equation}
U=\left(\frac{\partial W}{\partial \phi}\right)^2 - W^2.
\end{equation}
If one considers a single vector supermultiplet  then supersymmetry imposes that
\begin{equation}
W=\xi e^{\alpha \phi}
\end{equation}
where $\alpha=1/\sqrt 3,\ -1/\sqrt 12$, these values arising from the
parametrisation of the moduli space of the vector multiplets, and $\xi$ is
a characteristic scale related to the brane tension. 

Since supersymmetry is not observed in nature, one should incorporate 
supersymmetry breaking. A natural way to break supersymmetry is  
by coupling the bulk scalar field to brane fields fixed at their vevs.
This leads to
\begin{equation}
U_B=TW
\end{equation}
where $T=1$ is the supersymmetric case. Larger values of $T$
correspond to supersymmetry breaking effects with a positive energy
density on the brane.
We will analyse the dynamics of the coupled system comprising gravity,
the scalar field $\phi$ and matter on the brane.

The dynamics  of the brane world is specified by the four dimensional 
Einstein equations \cite{mennim}
\begin{equation}
\bar G_{ab}=-\frac{3V}{8}n_{ab} +\frac{U_B}{4}\tau_{ab}+\pi_{ab}
+\frac{1}{2}\nabla_a\phi\nabla_b\phi -\frac{5}{16}(\nabla\phi)^2 n_{ab}-E_{ab},
\end{equation}
where $\tau_{ab}$ is the matter energy momentum tensor,
$n_{ab}$ the induced metric on the brane and $E_{ab}$ the projected
Weyl tensor of the bulk onto the brane. It appears as an effective 
energy momentum tensor on the brane, called the Weyl fluid \cite{RScosmo}.

The tensor $\pi_{ab}$ is quadratic in the matter energy momentum tensor
\begin{equation}
\pi_{ab}=\frac{\tau}{12}\tau_{ab}-\frac{\tau_{ac}\tau_{b}^{c}}{4}+\frac{\tau_{ab}}{24}(
3\tau_{cd}\tau^{cd}-\tau^2)
\end{equation}
where $\tau=\tau^a_{~a}$,
The Bianchi identity $\nabla^a G_{ab}=0$ leads to the conservation equation
\begin{equation}
\nabla^a E_{ab}=\frac{\nabla^a
U_B}{4}\tau_{ab}+\nabla^a\pi_{ab}+\nabla^a P_{ab},
\label{cons}
\end{equation}
where the tensor $P_{ab}$ is defined by
\begin{equation}
P_{ab}=-\frac{3V}{8}n_{ab}+\frac{1}{2}\nabla_a\phi\nabla_b\phi -\frac{5}{16}(\nabla\phi)^2 n_{ab}
\end{equation}
This consistency equation (\ref{cons}) will allow us to follow the
time evolution of $E_{00}$ in the background and at the perturbative
level. This is crucial in order to define the time evolution of the
background and of the scalar field and matter perturbations.

The dynamics of the scalar field is specified by the Klein-Gordon
 equation which reads
\begin{equation}
\nabla^2
\phi+\frac{\tau}{6}\frac{\partial U_B}{\partial\phi}=\frac{\partial V}{\partial\phi}-\Delta\Phi_2
\end{equation}
where we have defined the loss parameter by
\begin{equation}
\Delta \Phi_2= \partial_n^2\phi \vert_0 -\frac{\partial U_B}{\partial \phi}\frac{\partial^2 U_B}{\partial \phi^2}
\end{equation}
and $\partial_n^2\phi\vert_0$ stands for the second normal derivative of the
scalar field at the brane location. The effective scalar potential is
defined by
\begin{equation}
V=\frac{U+(\frac{\partial U_B}{\partial \phi})^2-U_B^2}{2}.
\end{equation}
We derive these equations  in section 3.

In order to illuminate the role of $U_B$ further, we point out that 
the projected Klein-Gordon equation can be seen as an equation for 
the brane energy-momentum tensor
\begin{equation}
{\cal T}_{ab}=\nabla_a\phi\nabla_b\phi -\frac{1}{2}((\nabla\phi)^2 +2V) n_{ab},
\end{equation}
which reads
\begin{equation}
\nabla^a{\cal T}_{ab}= -\frac{\tau}{6}\nabla_b U_b-\Delta\Phi_2\nabla_b \phi 
\end{equation}
and can be derived from the four dimensional effective action for $\phi$
\begin{equation}
S=\int d^4 x \sqrt{-g_4}\left( (\nabla\phi)^2 +2V-\frac{\tau}{3}U_B 
- 2\Delta\Phi_2\phi \right),
\end{equation}
whenever $\Delta\Phi_2$ is constant. 
It is remarkable that $U_B$ plays the role of a dilaton coupled to
the trace of the matter energy-momentum tensor.

For the background cosmology the Klein-Gordon equation reduces to (see
section 3)
\begin{equation}
\ddot\phi + 4H \dot\phi -\frac{\tau}{6}U_B'=-V' +\Delta\Phi_2
\end{equation} 
where dot stands for the proper time derivative.
The bulk expansion rate is defined by
$4H\equiv \partial_{\tau}\ln \sqrt{-g}\vert_0$
evaluated on  the brane.

The background cosmology is characterized by its isotropy so we
consider that $E_{0i}=0, E_{ij}=0,\ i\ne j$. Moreover we assume that we obtain  a FRW induced
metric on the brane. Due to the tracelessness of $E_{ab}$ 
it is then sufficient to obtain the differential equation for $E_{00}$
\begin{equation}
\dot E_{00}+ 4H_B E_{00}=\partial_t\left( \frac{3}{16}\dot \phi^2 +
\frac{3}{8}V \right)+\frac{3}{2}H_B\dot\phi^2
+\frac{\dot U_B}{4}\rho,
\end{equation}
where $H_B$ is the brane expansion rate 
$3H_B\equiv \partial_{\tau}\ln \sqrt{-g_B}\vert_0$. This leads to 
\begin{equation}
E_{00}=\frac{1}{a^4}\int dt a^4 \left(\partial_t \left( \frac{3}{16}\dot \phi^2 +\frac{3}{8}V\right)
+\frac{3}{2}H_B\dot\phi^2+ \frac{\dot U_B}{4}\rho \right).
\end{equation}
Writing the bulk metric as
\begin{equation}
ds^2=e^{2A(z,t_b)}(-dt_b^2+dz^2)+e^{2B(z,t_b)}dx^2,
\end{equation}
the proper time on the brane is defined by $dt=e^{A(0,t_b)}dt_b$.
We can always choose
the boundary condition $A(0,t_b)=B(0,t_b)$ in such a way 
\begin{equation}
H=H_B.
\end{equation}
Together with the Klein-Gordon equation this leads to
\begin{equation}
E_{00}=\frac{1}{16 a^4}\int dt a^4 \left( \dot U_B ( 4\rho
-\tau)+6\Delta\Phi_2\dot\phi \right)
\end{equation}
In particular we find that conformal flatness is broken as soon as
$\phi$ becomes time-dependent and either matter is present on the
brane
or the energy loss parameter does not vanish.
Notice that a sufficient  condition for breaking conformal flatness is
that  Newton's constant becomes time-dependent. 
We will comment on this expression when discussing the various
cosmological eras.

Using
\begin{equation}
\bar G_{00}=3H_B^2 
\end{equation}
and after one integration by parts we obtain  the Friedmann equation
\begin{equation}
H_B^2=\frac{\rho^2}{36}+\frac{U_B}{12}\rho +\frac{1}{16 a^4}\int dt
\frac{da^4}{dt}(2V-\dot\phi^2)-\frac{1}{12 a^4}\int dt \frac{dU_B}{dt}\rho + \frac{C}{a^4}
\end{equation}
This equation has already been derived in \cite{BDII}. Notice that the
scalar field $\phi$ enters in the definition of the effective Newton's 
constant. The last three terms are a combination of the energy flow
onto or away from the brane and the changes of the pressure along the 
fifth dimension \cite{vdB}.

Using the conservation of matter
\begin{equation}
\nabla_a \tau^a_b=0
\end{equation}
we find that 
\begin{equation}
\dot \rho=-3H_B(\rho+p).
\end{equation}
This completes the description of the three equations determining the
brane cosmology, i.e.  the Klein-Gordon equation, the Friedmann
equation and the
conservation equation.

Notice that there are two   entities which depend on the bulk.
First of all there is the dark radiation term $C/a^4$ whose origin
springs from the possibility of black-hole
formation in the bulk. Then there is the loss parameter $\Delta\Phi_2$
which depends on the evolution of the scalar field in the bulk. It 
specifies the part of the evolution of $\phi$, which is not
constrained by considerations of the brane dynamics.

In the following we shall describe the case where the bulk theory is
$N=2$ supergravity with vector multiplets. When no matter is present
on the brane, the background cosmology can be explicitly solved.
In particular the metric is conformally flat implying that
\begin{equation}
C=0.
\label {C}
\end{equation}
Moreover the loss parameter vanishes explicitly
\begin{equation}
\Delta\Phi_2=0.
\label{p}
\end{equation}
A detailed analysis is presented in the appendix. 
In particular the last equation implies that the brane dynamics is closed.
When  matter density is present on the brane, we show in the appendix
that conformal flatness is not preserved. Nevertheless we assume that
the breaking of conformal flatness is small enough to allow one to
assume that (\ref{C}) and (\ref{p}) are still valid, both for the 
background as well as on the perturbative level. 

We now turn to the discussion of the background evolution. The
solutions given below have been obtained in \cite{BDII}, but
we need to present them in detail in order to discuss the evolution of
density perturbations in  section 4. 

\subsection{Radiation Dominated Eras}

The background cosmology can be solved in four different eras:
the high energy era, the radiation and matter dominated epochs and
finally
the supergravity era where the scalar field dynamics dominates.
We discuss these four eras in turn paying particular attention to the
matter-supergravity transition where we show that requiring
coincidence
now implies a fine-tuning of the supersymmetry breaking part of the
brane tension. 

First of all we consider the high energy regime where the
non-conventional
$\rho^2$ dominates, i.e. for energies higher than the brane tension.
In that case we can neglect the effective potential $V$ and find that 
\begin{equation}
a=a_0\left(\frac{t}{t_0}\right)^{1/4},
\end{equation}
while the scalar field behaves like
\begin{equation}
\phi=\phi_0+\beta\ln \left(\frac{t}{t_0}\right).
\end{equation}
In the following we will focus on the case $\beta=0$, i.e. a constant
scalar field, for which the projected Weyl
tensor vanishes altogether
\begin{equation}
E_{00}=0.
\end{equation}
Notice that Newton's constant does not vary in time in this case.

The usual radiation era is not modified by the presence of the scalar field
\begin{equation}
\phi=\phi_0
\end{equation}
and
\begin{equation}
a=a_r\left( \frac{t}{t_r}\right)^{1/2}.
\end{equation}
Newton's constant does not vary in time, while
\begin{equation}
E_{00}=0,
\end{equation}
as in the high energy regime. Note that the solution in the radiation
era in this brane world scenario is similar to the radiation era
solution found in Brans--Dicke theory, see \cite{Nordtvedt} and
\cite{Barrow}, for example. The field $\phi$ approaches quickly 
the attractor for which $\phi=$constant. 

\subsection{Matter Dominated Era}

The matter dominated era leads to a more interesting background
cosmology.
Let us first consider the pure sugra case where $T=1$. This is a good
approximation until coincidence where the potential energy of the
scalar field cannot be neglected anymore.
The solution to the evolution equations is
\begin{eqnarray}
\phi&=&\phi_0+\beta\ln\left(\frac{t}{t_e}\right) \\
a&=&a_e\left(\frac{t}{t_e}\right)^{\gamma}
\end{eqnarray}
where $\tau_e$ and $a_e$ are the time and scale factors at
radiation--matter equality. 
We are interested in the small $\alpha$ case as it leads to an accelerating universe
when no matter is present and  small time deviations
for Newton's constant.

For small $\alpha$ we get 
\begin{eqnarray}
\beta&=&- \frac{8}{15}\alpha\nonumber \\
\gamma&=&\frac{2}{3}-\frac{8}{45}\alpha^2.\nonumber\\
\end{eqnarray}
The projected Weyl tensor is given by 
\begin{equation}
E_{00}=-\frac{4\alpha^2}{3t^2},
\end{equation}
which decreases like $a^{-3}$ to leading order in $\alpha$.

In a phenomenological way we identify Newton's constant with the ratio
\begin{equation}
\frac{8\pi G_N(\tau)}{3}\equiv \frac{H^2}{\rho_m}.
\end{equation}
In terms of the red-shift $z$ this is 
\begin{equation}
\frac{G_N(z)}{G_N(z_e)}=\left(\frac{z+1}{z_e+1}\right)^{4\alpha^2/5}.
\end{equation}
For the supergravity case with $\alpha^2=1/12$ the exponent is $1/15$.
As $z_e\sim 10^3$ this leads to a decrease   by $37\%$ since equality.

Notice that the Newton constant starts to decrease only from the time of
 matter and radiation equality and is
strictly constant during the radiation dominated era. Nucleosynthesis 
constrains the variation to be less than $20\%$. 
However, in our model we would expect the couplings to standard model
particles to also vary in a similar manner. This could lead to a variation
in, for example, the proton and neutron masses since these arise from Yukawa
couplings in the standard model. We note that many of the 
tests for the variation of the Newton constant assume all other masses
and couplings are constant \cite{wetterich}; it is possible that our 
supergravity variation would evade detection. 

\subsection{Supergravity Era}

After coincidence matter does not dominate anymore; this is the
supergravity era dominated by the scalar field dynamics.
Let us review it briefly. Consider first the pure sugra case.
It is easy to see that the potential vanishes
\begin{equation}
V_{SUGRA}=0
\end{equation}
leading to a static universe with
\begin{eqnarray}
\phi&=&-\frac{1}{\alpha}\ln (1-\alpha^2\xi\vert y\vert)\nonumber \\
a&=&(1-\alpha^2\xi\vert y\vert)^{1/4\alpha^2}\nonumber \\
\end{eqnarray}
where we have defined $dy=adx_5$.
This is a flat solution corresponding to a vanishing cosmological constant on the brane-world.

As soon as supergravity is broken on the brane $T\ne 1$ the static
solution is not valid anymore.
The new four dimensional potential becomes
\begin{equation}
V=\frac{(T^2-1)}{2}\left(W^2-\left(\frac{\partial W}{\partial \phi}\right)^2\right).
\end{equation}

The  time dependent background  is obtained from the static solution
by going to conformal coordinates 
\begin{equation}
ds^2=a^2(u)(-d\eta^2+du^2+dx^idx_i)
\end{equation}
and performing a boost along the $u$ axis
\begin{eqnarray}
a(u,\eta)&=& a\left(u+h\eta,\frac{\xi}{\sqrt {1-h^2}}\right)\nonumber \\
\phi(u,\eta)&=& \phi\left(u+h\eta,\frac{\xi}{\sqrt {1-h^2}}\right)\nonumber \\ 
\end{eqnarray}
where we have displayed the explicit $\xi$ dependence.
Now for 
\begin{equation}
h=\pm \frac{\sqrt{T^2-1}}{T}
\end{equation} 
we find that the Friedmann equation is fulfilled. Similarly the Klein-Gordon equation is satisfied.
Moreover we find that 
\begin{equation}
E_{00}=0,
\end{equation}
as the bulk metric is conformally flat.

The resulting universe is characterized by the scale factor in cosmic time
\begin{equation}
a(t)=\frac{1}{\sqrt T}\left(1-\frac{t}{t_0}\right)^{1/3+1/6\alpha^2}
\end{equation}
with $t_0=\frac{2}{3\alpha^2}\frac{1}{hT^{3/2}\xi}$. 
The scale factor 
corresponds to a solution of the four dimensional FRW equations with an
acceleration parameter 
\begin{equation}
q_0=\frac{6\alpha^2}{1+2\alpha^2}-1
\end{equation}
and an equation of state
\begin{equation}
\omega_{SUGRA}=-1+\frac{4\alpha^2}{1+2\alpha^2}
\end{equation}
which never violates the dominant energy condition.
The solution with $\alpha=-\frac{1}{\sqrt{ 12}}$ is accelerating. 
In particular we find that 
\begin{equation}
q_0=-\frac{4}{7}
\end{equation}
and
for the equation of state
\begin{equation}
\omega_{SUGRA}=-\frac{5}{7}.
\end{equation}
This is within the experimental ball-park.

\subsection{The Matter-Supergravity Transition}

Let us now investigate the transition between the matter dominated and
supergravity eras.
If we denote by $H_{SUGRA}$ the Hubble parameter derived in the pure 
supergravity case, then 
the Friedmann equation in the broken supergravity case is
\begin{equation}
H^2=H^2_{SUGRA}+\frac{V}{8},
\end{equation}
where we have used the fact that $\phi$ varies slowly compared to $\alpha$. 
The evolution coincides with the one obtained from unbroken  supergravity
as long as the contribution from the potential does not dominate.
In the radiation dominated era this requires
\begin{equation}
\frac{T^2-1}{T}\frac{W}{2\kappa^2_5} \ll \frac{2}{3}\frac{\rho_e}{1-\alpha^2},
\end{equation}
where $\rho_e$ is the matter density at equality. This implies that the left-hand side
is much smaller that $10^{-39}$ GeV$^4$.
Let us now denote the supersymmetric brane tension by 
\begin{equation}
M_S^4=\frac{3W}{2\kappa_5^2}
\end{equation}
and the supersymmetry breaking contribution
\begin{equation}
M_{BS}^4=(T-1)M_S^4.
\end{equation}
We find that 
\begin{equation}
M^4_{BS} \ll \frac{\rho_e}{1-\alpha^2}.
\end{equation}
Now this is an extreme fine-tuning of the non-supersymmetric contribution to the brane tension.

In the matter dominated era the supergravity Hubble parameter decreases faster than the 
potential contribution.
Coincidence between the matter dominated supergravity contribution
$H^2_{SUGRA}$ and the potential energy occurs at $z_c$ 
\begin{equation}
M^4_{BS}=\frac{1}{1-\alpha^2}\left(\frac{z_c+1}{z_e+1}\right)^{3+\alpha\beta/\gamma}\rho_e.
\end{equation}
Imposing that coincidence has occurred only recently leads to 
\begin{equation}
M^4_{BS}\approx \rho_c,
\end{equation}
where $\rho_c$ is the critical density. This is the usual extreme
fine-tuning of the cosmological constant. Indeed it 
specifies that the energy density received by the brane-world from the 
non-supersymmetric sources, e.g. radiative corrections and phase 
transitions, cannot exceed the critical energy density of the universe. 

\section{Cosmological Perturbations using the Fluid Flow Approach}
We now turn to the discussion of cosmological perturbations. There are 
different effects which will influence the evolution of perturbations:

\begin{itemize}
\item The evolution of the gravitational constant in the matter era
changes the evolution of the background: In the matter era, the 
gravitational constant decreases  and furthermore the universe is expanding 
slower (up to order $\alpha^2$) than in the FRW matter dominated era 
in general relativity. 

\item Perturbations in the scalar field are the source of matter
fluctuations  and vice versa. 

\item Perturbations in the  projected Weyl tensor act as sources for
the scalar and matter perturbations and vice versa. 

\end{itemize}

These effects will change the growth of perturbations compared to
normal 4D cosmological models or the Randall--Sundrum model. 
In particular it should be noted that our formalism allows us to 
treat the Randall--Sundrum cosmology.
Indeed by putting $\alpha=0$, $T=1$ and neglecting the scalar field
contribution we obtain  the Randall-Sundrum case with a flat boundary
brane while putting $T \ne 1$ leads to a de Sitter boundary brane.  

While discussing perturbations, we  use the fluid--flow approach (see 
\cite{Hawking}, \cite{Lyth} and \cite{Liddle}) rather than the 
metric--based approach \cite{robert}. The main
difference is that all perturbation variables are expressed in terms
of fluid--quantities, rather than metric--variables. For our purpose,
i.e. discussing the evolution of the density contrast $\delta = \delta
\rho/\rho$ of the dominant fluid at each epoch, this approach is
simpler in order to obtain the evolution equations. To do so, we 
need to derive the Raychaudhuri equation and the Klein--Gordon 
equation in the comoving frame. 

\subsection{The Raychaudhuri and Klein-Gordon Equation}
On the brane $\bf\cal{B}$, see figure 1, the matter energy-momentum tensor is conserved. Denoting by
$n$ the normal vector to the brane and defining the induced metric by
\begin{equation}
n_{ab}=g_{ab}-n_an_b
\end{equation}
such that $n^2=1$ and $n_{ab}n^b=0$, the conservation equation reads
\begin{equation}
\nabla_a \tau^{ab}=0
\label{con}
\end{equation}
where $\nabla_a=n_a^bD_b$ is the brane covariant derivative and
\begin{equation}
\tau_{ab}= (\rho+p)u_au_b -pn_{ab}
\end{equation}
is the energy-momentum tensor. Notice that the vector $u_a$ is the
velocity field of the brane matter and thus must be orthogonal to
$n$ and satisfies $u^2=-1$. 
In addition we have  the usual decomposition in terms of the shear $\sigma_{ab}$,
the helicity $\omega_{ab}$
and the expansion rate
\begin{equation}
\nabla_c u^d=-u_c\dot u^d +H_B u_c^d +\sigma^d_c+ \omega_c^d
\label{dec}
\end{equation}
where we set  $\sigma_{ab}=0,\ \omega_{ab}=0$ later on. 

Using the relations 
\begin{equation}
\nabla_a u^a=3H_B,\ \dot u_a=u^b\nabla_b u_a
\end{equation}
one derives from (\ref{con})
\begin{equation}
\dot \rho=-3H_B(\rho+p).
\end{equation}
Notice that no matter leaks out of the brane. 
Defining
\begin{equation}
\bar \nabla_a\equiv u_{ab}\nabla^b=\nabla_a+u_au^bD_b,
\end{equation}
which is nothing but the spatial covariant derivative on the brane, 
one obtains
\begin{equation}
\dot u_a=-\frac{\bar \nabla_a p}{\rho+p},
\end{equation}
whose divergence $\nabla_a \dot u^a$ leads to the Raychaudhuri equation
\begin{equation}
3\dot H_B+ 3H_B^2 =-R_{00} -\bar \nabla_a(\frac{\bar \nabla_a p}{\rho+p})
\end{equation}
We have neglected $\sigma_{ab}$ and $\omega_{ab}=0$ here. 

\FIGURE{
\epsfxsize=10.cm
$$\epsfbox{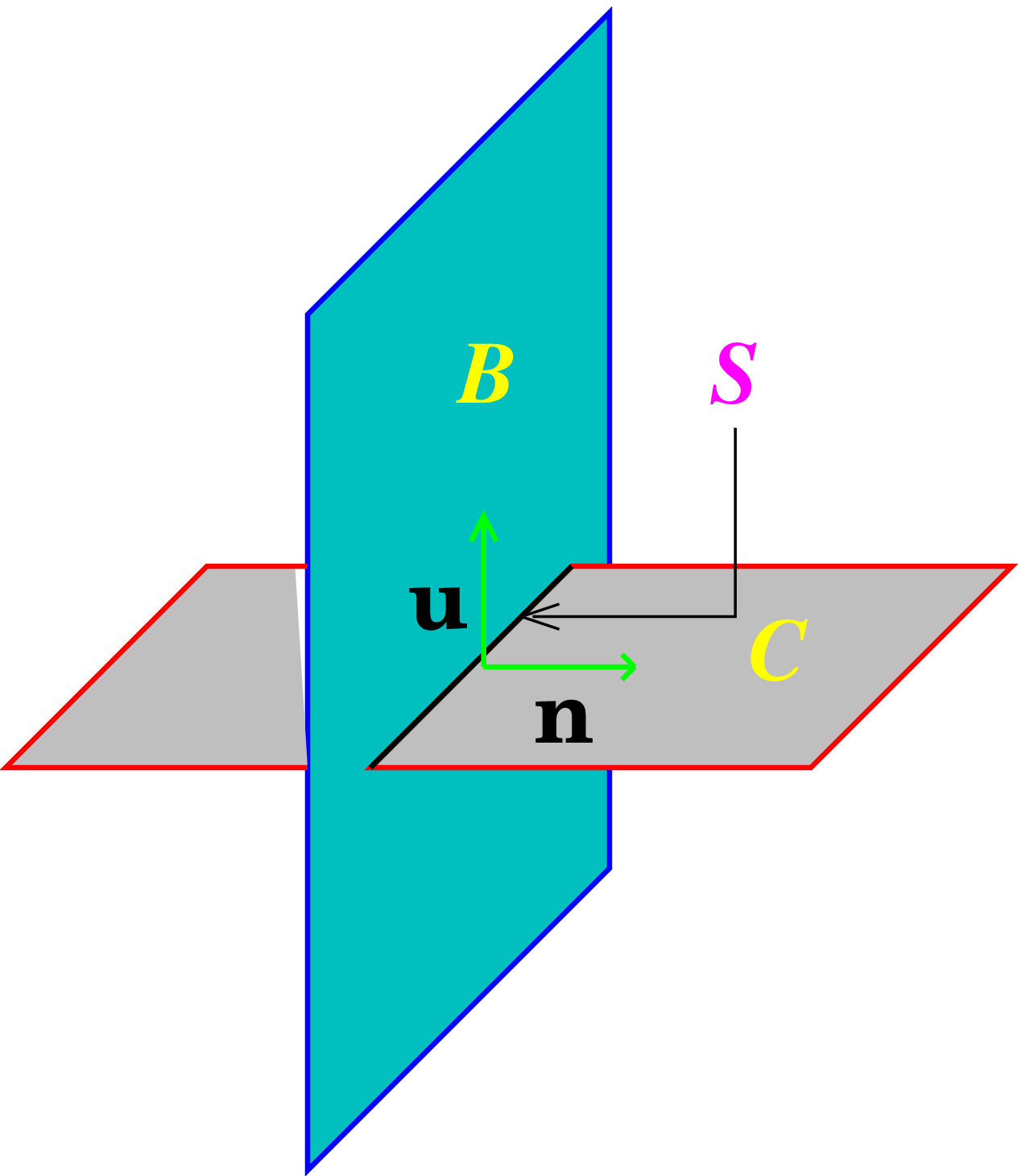}$$
\caption{ The brane-world ${\cal B}$ is perpendicular to the normal vector
$n$. The four dimensional hypersurface ${\cal C}$ is orthogonal to the
velocity vector $u$. The surface $S=C\cap B$ is a set of comoving 
observers following matter on the brane. Note that the perturbed brane
is not necessarily located at $y=$constant. The different metrics 
in the text are as follows: $g_{ab}$ is the full five--dimensional 
metric, $h_{ab}$ is the induced metric on the brane ${\cal B}$, 
$u_{ab}$ is the induced metric on ${\cal C}$ and $n_{ab}$ is the 
induced metric on the comoving hypersurface ${\cal S}$.}
}

In order to have a closed system of differential equations
for the perturbations we need  the Klein-Gordon equation in the comoving
gauge. 
The metric on a hypersurface $\bf\cal{C}$ orthogonal to $u$, see
figure 1, is given by
\begin{equation}
h_{ab}=g_{ab}+u_au_b,
\end{equation}
where $u^2=-1$ and $ h_{ab}u^b=0$.
Defining the covariant  derivative on the hypersurface $\bf{\cal C}$ by 
$\bar D_a=h_{ab}D^b$ and using $ D_au^a=4H$ one obtains
\begin{equation}
D_aD^a=\bar D_a \bar D^a
-4H u_c D^c -u_au_b D^aD^b.
\end{equation}
In the comoving gauge we have $u_0=-1,u_i=0$ leading to
\begin{equation}
D^2\phi=\bar D^2 \phi - 4H \dot \phi -\ddot \phi.
\end{equation}
We can now evaluate the Laplacian $\bar D^2$ in terms of the
Laplacian
on the comoving surface $\bf {\cal S}$ (see figure 1) orthogonal to $u$. 
Notice that 
$\bar \nabla_a =  n_{ab}\bar D^b$.
Expanding $\bar \nabla ^2$ leads to
\begin{equation}
\bar D^2=\bar \nabla^2 +(K+u^au^bK_{ab})n_c\bar D^c +n_an_b\bar
D^a\bar D^b
\end{equation}
Now $K_{ab}u^au^b=u^aD_a(u \cdot n)-\dot u_a n^a$, where
\begin{equation}
K_{ab}=D_an_b
\end{equation}
is the extrinsic curvature tensor.
Using $(u \cdot n)=0$ and $n^a\bar\nabla_a p=0$ we find that
\begin{equation}
K_{ab}u^au^b=0.
\end{equation}
The junction conditions lead to
\begin{equation}
K=\frac{\cal \tau }{6}- U_B,
\end{equation}
with $\tau=-\rho+3p$.
Finally we can read off the Klein-Gordon equation
\begin{equation}
\ddot \phi +4H\dot\phi -\phi''-\left(\frac{\tau}{6}-U_B\right)\phi'-\bar \nabla^2
\phi=-\frac{1}{2}\frac{\partial U}{\partial \phi}
\end{equation}
where prime stands for the normal derivative.
Now the junction conditions lead to
\begin{equation}
\phi'=\frac{\partial U_B}{\partial \phi}.
\end{equation}
We find that the Klein-Gordon equation reduces to
\begin{equation}
\ddot \phi +4H\dot\phi +\frac{1}{6}\frac{\partial U_B}{\partial \phi}\left(\rho-3p\right)-\bar \nabla^2\phi 
=-\frac{\partial V}{\partial \phi}+ \Delta\Phi_2.
\end{equation}
Notice that it involves only brane derivatives when the loss parameter $\Delta\Phi_2=0$. 
We now turn to the evolution equation for $E_{00}$ which enters in
the evaluation of $R_{00}$.

\subsection{The Evolution Equation for $E_{\mu\nu}$}

Let us consider  the  hypersurface ${\cal S}$ on the brane with induced metric
$u_{ab}$ and orthogonal to the velocity vector $u^a$ such that 
 $u_{ab}u^a = 0$ and $u^a u_a = -1$. The
 induced metric is given by $u_{ab} = n_{ab} 
+ u_a u_b$, where $n_{ab}$ is the brane metric. Let $\nabla_a$ 
be the brane  covariant derivative and $\bar{\nabla}_a$ the 
covariant derivative with respect to $u_{ab}$.

Consider now $\nabla^a E_{ab}$:
\begin{eqnarray}\label{one}
\nabla^a E_{ab} = u^{ac}\nabla_c E_{ab} - u^a u^c \nabla_c E_{ab}.
\end{eqnarray}
The last term  can be written as
\begin{equation}
u^a u^c \nabla_c E_{ab} = u^c \nabla_c (u^a E_{ab}) - \dot u^a E_{ab}.
\end{equation}
The first term in equation (\ref{one}) can be rewritten as
\begin{eqnarray}
u^{ac}\nabla_c E_{ab} &=& \nabla_c (u^{ac}E_{ab}) - E_{ab} (\nabla_c
u^{ac}) \nonumber \\
&=& \nabla_c (u^{ac}E_{ab}) - E_{ab} \left(\dot u^a + 3 H_B u^a
\right). 
\end{eqnarray} 
As a next step we consider 
\begin{eqnarray}\label{two}
\nabla_c (u^{ac}E_{ab}) &=& \nabla_c (u^{ac}n^{d}_{~b} E_{ad})
\nonumber \\
&=& \nabla_c (u^{ac}u^{d}_{~b} E_{ad}) - \nabla_c (u^{ac}u^{d}u_{b} E_{ad})
\nonumber \\
&=& \nabla_c (u^{ac}u^{d}_{~b} E_{ad}) - u^d \nabla_c(u^{ac} u_b E_{ad})
- (\nabla_c u^d)u^{ac}E_{ad}u_b.
\end{eqnarray}
Using (\ref{dec})
\begin{equation}
(\nabla_c u^d)u^{ac}E_{ad}u_b = H_B u^{ad}E_{ad}u_b 
+ (\sigma_c^{~d} + \omega_c^{~d})u^{ac}E_{ad}u_b.
\end{equation}
Let us further define the projected tensor
\begin{equation}
\bar E_{ab} = u_a^{~c} u_b^{~d}E_{cd}
\end{equation}
Then
\begin{equation}
\nabla_c \bar E^{c}_{~d} = \bar \nabla_c \bar E^{c}_{~d}
- \dot u_c \bar E^{c}_{~d}.
\end{equation}
So, from eq. (\ref{two}) we get
\begin{eqnarray}
\nabla_c (u^{ac}E_{ab}) &=& \bar \nabla_c \bar E^{c}_{~b} 
- \dot u_c \tilde E^{c}_{~b} - u^d \nabla_c (u^{ac}E_{ac}u_b) 
- H_B u^{ad}E_{ad}u_b \nonumber \\
&-& (\sigma_c^{~d} + \omega_c^{~d})u^{ac}E_{ad}u_b.
\end{eqnarray}
Collecting everything and defining  $\bar E = u^{ab}E_{ab}$ 
,the spatial trace,  
we end up with
\begin{eqnarray}
\nabla^a E_{ab} &=& \bar \nabla_c \bar E^c_{~b} 
- \dot u_c \bar E^c_{~b} - u^d \nabla_c (u^{ac}E_{ad}u_b) 
-(u^a E_{ab})^{.} - H_B \bar E u_b - 3H_B u^a E_{ab} \nonumber \\
&-& (\sigma_c^{~d} + \omega_c^{~d})u^{ac}E_{ad}u_b.
\end{eqnarray}

Let us now specialize to the comoving gauge, i.e. the surface 
${\cal S}$ is a comoving one and $u^0=1,\ u^i=0$.
Notice that the first term of the last expression vanishes. 
We obtain
\begin{equation}
\nabla^aE_{a0}=\bar \nabla_c E^c_0 -4H_B E_{00}-\dot E_{00} -
\dot u_a E^a_0
+(\sigma_c^d+\omega_c^d)u^{ac}E_{ad}
\end{equation}
where we have used the tracelessness of $E_{ab}$ to get $\bar E=E_{00}$.
For the background we retrieve the previous expression as $\dot u_a=0$
and $E^i_0=0$ and we explicitly assume that $\sigma=0, \ \omega=0$ for 
the background. 

We need further to to evaluate
\begin{equation}
\nabla^a \pi_{ab}=\bar \nabla^c \pi_{c0}-3H_B \pi_{00}-H_B \bar \pi
-\dot u_a \pi^a_0 -\dot \pi_{00}
\end{equation}
which simplifies to
\begin{equation}
\nabla^a \pi_{ab}=\bar \nabla^c \pi_{c0}
-\dot u_a \pi^a_0 
\end{equation}
and finally we need
\begin{equation}  
\nabla^a P_{a0}=\bar \nabla^c P_{c0}-3H_B P_{00} -H_{B}\bar P
-\dot u_a P^a_0 -\dot P_{00}
\end{equation}
Collecting all these ingredients leads to the consistency equation
(\ref{cons}) expressed in the comoving frame.

\subsection{The Perturbed Dynamics}
We have now obtained all the necessary equations  in order to derive the 
perturbation equations up to linear order. In doing so we decompose 
all fluid quantities into an average (over the comoving hypersurface)
of the quantity plus a perturbation, i.e. $\rho(x,t) = \rho_b(t) +
\delta\rho (x,t)$, and similar for the pressure and the expansion
rate $H$. We insert these expressions for all quantities into the 
equations we obtained above and subtract the average.

We begin with the Raychaudhuri equation and energy conservation
equation evaluated in a comoving basis. The Raychaudhuri equation 
reads
\begin{equation}\label{Raychaudhuri}
3 \dot H + 3H^2 = -\frac{\nabla^2 p}{\rho + p} - R_{00}.
\end{equation}
In this expression $H$ is the expansion rate, $p$ the pressure 
and $\rho$ the energy density. All quantites are measured with
respect to a comoving observer\footnote{This observer is, of course, 
confined onto the brane world.}. $R_{00}$ is the time--time component of
the Ricci tensor. The dot represents the time--derivative with respect
of the comoving observer (i.e. proper time). The Raychaudhuri 
equation expresses just the behaviour of matter under the 
influence of geometry and is independent of the field equation. 
In particular, for matter on the brane, it is the same as in the 
usual 4D case. The five--dimensional character of the spacetime 
enters only through  $R_{00}$. For the FRW metric,
$R_{00}$ is
\begin{equation}\label{Ricci}
R_{00} = -\frac{3}{8}V + \frac{U_b}{8}\left( \rho + 3 p \right) + 
\frac{1}{12}\rho\left(2\rho + 3p \right) - \frac{7}{16}\dot\phi^2
- E_{00}.
\end{equation}
The proper time is not a unique label of the comoving hypersurfaces,
because it can vary in space. Therefore, we need to transform to 
coordinate time, which labels these comoving hypersurfaces. This 
transformation from proper time to coordinate time is given by 
(see \cite{Lyth} or \cite{Liddle})
\begin{equation}\label{timetrans}
\frac{dt_{pr}}{dt} = 1 - \frac{\delta p}{\rho + p}.
\end{equation}
This gives the variation of proper time due to the perturbations.
Using the conservation equation we obtain
\begin{equation}
\delta H=H\frac{\omega}{1+\omega}\delta -\frac{\dot \delta}{3(1+\omega)}.
\end{equation}
where we have defined the density contrast
\begin{equation}
\delta=\frac{\delta \rho}{\rho}.
\end{equation}
In the rest of the paper we denote by dot the coordinate time
derivative of the perturbations.
In this paper we concentrate on the case of $w = p / \rho = {\rm
const.}$ and $c_s^2 = w$. 
We  deduce that
\begin{equation}
\delta\dot H= \frac{\omega}{1+\omega}(H\dot \delta+\dot H \delta) -\frac{\ddot \delta}{3(1+\omega)},
\end{equation} 
which  combined with the perturbed Raychaudhuri equation
\begin{equation}\label{Raychaudperturbed}
\delta \dot H = \frac{\delta p}{\rho + p}\dot H - 2H\delta H + 
\frac{1}{3}\frac{k^2}{a^2}\frac{\delta p}{\rho + p} -
\frac{1}{3}\delta R_{00}
\end{equation}
leads to an equation for the density constrast 
\begin{equation}
\ddot\delta +(2-3\omega)H\dot\delta-6\omega(H^2+\dot
H)\delta=(1+\omega)\delta R_{00}-\omega \frac{k^2}{a^2}\delta 
\end{equation}
where we have used $\bar \nabla^2= k^2/a^2$.
Notice that the left-hand side coincide with the usual four
dimensional
expression. The new physical ingredients all spring from $\delta R_{00}$.
As the Friedmann
equation and its time-derivative are given by complicated expressions, 
we do not derive a general equation for the evolution of $\delta$ but
rather analyse each regime separately and derive the corresponding 
equation. 

In the same fashion, we derive the perturbed Klein-Gordon equation, 
using eq. (\ref{timetrans}):
\begin{eqnarray}\label{scalarevolv}
(\delta \phi)^{..} &+& 4H (\delta \phi)^{.} 
+ \left[ \frac{k^2}{a^2} + \frac{\rho}{6}\left( 1 - 3w
\right)\frac{\partial^2 U_B}{\partial\phi^2}
+ \frac{\partial^2 V}{\partial\phi^2} \right] \delta \phi \nonumber \\
&=& \frac{c_s^2}{1+\omega} \left[ \dot \phi\dot \delta +
\delta(\ddot \phi +4H\dot\phi)\right] + \frac{1}{6}\frac{\partial
U_B}{\partial \phi}\left[ 3 \delta
p - \delta \rho \right].
\end{eqnarray}
where the coupling between the matter and scalar perturbations is explicit.

Finally we need the perturbed version of the $E_{00}$ consistency
equation. 
From the equations we derived in the previous section we obtain  
\begin{equation}
\delta (\nabla^a E_{a0})=\bar \nabla^c \delta E_{c0}-4\delta H_B \dot E_{00}
-4H_B \delta E_{00}-\delta \dot E_{00}-\frac{\delta p}{p+\rho}\dot E_{00},
\end{equation}
where the shear term vanishes automatically at first order. 
Similarly we have 
\begin{equation}
\nabla^a \delta \pi_{a0}=\bar \nabla^c \delta \pi_{c0}
-\dot u_a \pi^a_0 
\end{equation}
and 
\begin{equation}  
\nabla^a \delta P_{a0}=\bar \nabla^c \delta P_{c0}-3\delta H_B
P_{00}-3H_B\delta P_{00} -\delta H_{B}\bar P-H_B\delta \bar P
-\dot u_a P^a_0 - (\delta P_{00})^.
\end{equation}
This allows us to write down the necessary perturbed equation. 

Notice that the  components of $E_{a0}$ play a role in the perturbed
dynamics. As can be seen, they are not constrained by the brane
dynamics
and lead to a direct influence of the bulk perturbations on the brane
perturbations.
For this reason the perturbed dynamics is not closed on the brane.
In the rest of this paper we will only deal with long wave-length
phenomena where the dynamics is closed.

\section{Time Evolution of Cosmological Perturbations}
In this section we discuss some solutions to the perturbed dynamics. 
The equations are very difficult to solve in general so we 
restrict  ourselves to the different cosmological eras. We start in the 
high--energy regime, assuming the universe to be dominated by 
relativistic particles. Then we discuss the normal (low--energy)
radiation dominated epoch, followed by the matter dominated epoch. 
Because the system is not closed on the brane for large $k$ (small 
wavelength), we present solutions only for the $k\rightarrow 0$ limit.
This limit enables us to deduce the effects of the bulk scalar field 
and $E_{\mu\nu}$ on density perturbations.  

In particular we can write  the perturbed $E_{\mu\nu}$ equation in
that limit as
\begin{eqnarray}
&\ &\delta \dot E_{00}+\frac{\delta p}{\rho+p}\dot E_{00}+4H_B\delta E_{00}
+4\delta H_B E_{00}=
-\frac{\delta \dot U_B}{4}\rho -\frac{\delta p}{\rho+p}\frac{\dot U_B\rho}{4}
-\frac{\dot U_B\rho }{4}\delta+ \delta \dot P_{00}\nonumber\\
&\ &+\frac{\delta P}{\rho+p}\dot P_{00}+
3H_B\delta P_{00}+3\delta H_B P_{00}
 +H_B\delta \bar P + \delta
H_B \bar P\nonumber\\
\end{eqnarray}
Explicitly we need
\begin{equation}
P_{00}=\frac{3V}{8}+\frac{3\dot\phi^2}{16},\ \bar P= -\frac{9V}{16}+\frac{15\dot\phi^2}{16}
\end{equation}
The perturbed quantities are then
\begin{equation}
\delta \dot U_B= U_B'\delta\dot \phi
\end{equation}
and 
\begin{equation}
\delta P_{00}=\frac{3}{8}\dot \phi\delta\dot\phi
\end{equation}
as we neglect the potential in the different eras.
This equation will be made more explicit in each era.

As a first step we will first analyse the Randall-Sundum scenario with
a flat boundary brane. The differential equation for the
density contrast is of third order leading to the appearance of three
modes. This is to be compared to the usual two FRW cosmology modes
derived from four dimensional general relativity. 
As already mentioned two of the modes will coincide with FRW cosmology
while the third mode is entirely due to perturbations in 
the Weyl tensor signalling a breaking of conformal invariance in the bulk.

\subsection{The Randall--Sundrum scenario}
\subsubsection{The high energy era}
In the high energy regime, the equations for $\delta E_{00}$ and 
$\delta$ read
\begin{eqnarray}
\ddot \delta + H \dot\delta - 18 H^2 \delta &=& -\frac{4}{3}\delta E_{00}, \\
(\delta E_{00})^{.} + 4H(\delta E_{00}) &=& 0.
\end{eqnarray}
This leads to 
\begin{equation}
\delta E_{00} = \frac{\delta E_0}{t}.
\end{equation}
The overall solution of the equation for the density contrast is
a sum of the solution of the homogeneous equation as well as the 
general solution. It can easily be found as
\begin{equation}
\delta = \delta_0 t^{3/2} + \delta_1 t^{-3/4} 
+ \frac{16}{51}\delta E_0 t.
\end{equation}

\subsubsection{The low--energy radiation era}
In this era, the equations for $\delta E_{00}$ and 
$\delta$ read
\begin{eqnarray}
\ddot \delta + H \dot\delta - 2 H^2 \delta &=& -\frac{4}{3}\delta E_{00}, \\
(\delta E_{00})^{.} + 4H(\delta E_{00})&=& 0.
\end{eqnarray}
Again, the solution to the last equation is
\begin{equation}
\delta E_{00} = \frac{\delta E_0}{t^2}.
\end{equation}
The full solution to the first equation is found to 
be
\begin{equation}
\delta = \delta_0 t + \delta_1 t^{-1/2} + \frac{8}{3}\delta E_0.
\end{equation}
We do, therefore, find the normal growing and decaying modes  in this regime and 
a constant mode, which is absent in FRW cosmology. 

\subsubsection{The matter dominated epoch}
Finally, in the matter dominated epoch the equation for 
$\delta E_{00}$ and $\delta$ read
\begin{eqnarray}
\ddot \delta + 2 H \dot\delta - \frac{3}{2} H^2 \delta &=& - \delta E_{00}, \\
(\delta E_{00})^{.} + 4H(\delta E_{00}) &=& 0.
\end{eqnarray}
The solution to the last equation is
\begin{equation}
\delta E_{00} = \delta E_0 t^{-8/3},
\end{equation}
and the solution to the first equation can then found to be
\begin{equation}
\delta = \delta t^{2/3} + \delta_1 t^{-1} 
+ \frac{9}{4}\delta E_0 t^{-2/3}.
\end{equation}
We recover thus the normal growing and decaying mode as well as
a mode sourced by $\delta E_{00}$, which is decaying rapidly.

We now turn to the case with a scalar field in the bulk.
The previous Randall-Sundrum modes will be modified in two ways.
First of all the brane potential will lead to slight deviations
of the mode exponents similar to the modification
of the scale factor exponent in the matter era. Then
there will also appear new modes due to the fluctuations
of the scalar field governed by the Klein-Gordon equation.
We will now analyse each of the different regimes in turn. 

\subsection{Effects of the bulk scalar field on cosmological perturbations}

\subsubsection{The high--energy regime}
In this regime, the terms which are quadratic in $\rho$ and $p$
dominate both in the background and in the perturbation equations. 
We assume, that relativistic particles dominate the expansion, i.e.
$p=\rho/3$. 

Considering  the large wavelength limit is sufficient  
because all cosmologically relevant scales are 
far outside the horizon. The perturbation equations in 
this regime are
\begin{eqnarray}
&\ &\ddot \delta + H \dot\delta = 18 H^2 \delta - \frac{4}{3}\delta
E_{00}, \\
&\ &(\delta \phi)^{..} + 4H (\delta \phi)^{.} = 0, \\
&\ &(\delta E_{00})^{.} + 4H \delta E_{00} = - \frac{\alpha U_B
\rho}{4}(\delta \phi)^{.}
\end{eqnarray}
where we have assumed that $3H (\delta\phi)^. \gg V\delta
\phi$. 

To obtain the solutions to these equations, we first consider the 
Klein--Gordon equation. The solutions are  
\begin{equation}
\delta \phi = \delta \phi_0 + \delta \phi_1 \ln t,
\end{equation}
where $\delta \phi_0$ is  a constant mode.
We can now find $\delta E_{00}$.
\begin{equation}
\delta E_{00} = \frac{\delta E_0}{t}  - \frac{3\alpha U_B}{8}
\delta\phi_1\frac{\ln t}{t}
\end{equation}
Notice that the solution comprises two parts. The $1/t$ mode is a  solution of the homogeneous equation while the
logarithmic 
mode solves the complete equation. In the following we will always
find that the solutions are expressible as a sum of homogeneous modes
and modes solving the complete differential equations. 

We can now deduce the density constrast
\begin{equation}
\delta =\delta_0 t^{3/2} +\delta_1 t^{-3/4}
+\delta_2(t)
\end{equation}
where the complete solution reads
\begin{equation}
\delta_2(t)=-\frac{4}{3}t^{3/2}\int^t dt'(t')^{-13/4}\int^{t'} dt''
\delta E_{00}(t'')^{7/4}
\end{equation}
In the long time regime, we focus on the leading growing mode,
obtained by approximating
\begin{equation}
\delta_2(t)=O((\ln t) t)
\end{equation}
This implies
that we find two leading growing modes in $t^{3/2}$ and $(\ln t) t$.
Notice that the logarithmic mode is triggered by the scalar
fluctuation, $\delta\phi_1$, and is therefore absent in the
Randall-Sundrum case. 
Moreover for very long times we see that the leading mode increases
like $t^{3/2}$, which is much larger than in FRW
cosmology. Interestingly, this growth of fluctuations is anomalous, 
in the sense that the exponent of the growing mode is larger than 
one. The reason for this is the source term ($18 H^2 \delta)$ which 
is much larger than the normal $4\pi G \rho_m \delta$. In addition, 
there is the contribution of bulk gravity. In this regime we cannot 
expect the normal behaviour, because gravity is simply not four--dimensional. 

\subsubsection{Radiation domination}

In this regime the Newton's constant is not varying in time. 
Moreover matter does not appear in the Klein-Gordon equation. 
This leads to the  perturbation equations for the scalar field and the density 
contrast 
\begin{eqnarray}
(\delta \phi)^{..} + 4H(\delta \phi)^{.} &=& 0, \\
\ddot \delta + H \dot\delta - 2H^2 \delta &=& 4\alpha H^2 
\delta \phi - \frac{4}{3}\delta E_{00},\\
(\delta E_{00})^{.} + 4H \delta E_{00} &=& -3\alpha H^2 (\delta \phi)^{.}.
\end{eqnarray}
As before we find homogeneous and complete solutions. More specifically 
\begin{equation}
\delta \phi = \delta \phi_0 + \delta \phi_{-} t^{-1},
\end{equation}
leading to the perturbed $\delta E_{00}$
\begin{equation}
\delta E_{00} = \frac{\delta E_0}{t^2} + \frac{3\alpha}{4}\delta\phi_{-}
\frac{\ln t}{t^2}.
\end{equation}
Notice that the first mode springs from the homogeneous equation
and
the last one from the complete equation.  
The density contrast can then be deduced:
\begin{equation}
\delta =\delta_{1}t +\delta_{-1/2}t^{-1/2}
+t\int^tdt'(t')^{-5/2}\int^{t'}\left(
\frac{\alpha}{(t'')^{1/2}}\delta\phi
-\frac{4}{3}(t'')^{3/2}\delta E_{00}\right)
\end{equation}
There are two homogeneous modes as in FRW cosmology.
New contributions emerge from the scalar field. In particular there
is a growing mode in $O(\delta\phi_-\ln t)$ which is triggered by the
decreasing scalar mode in $\delta\phi_-/t$.

\subsubsection{Matter domination}

In this regime Newton's constant varies in time as the background
scalar field is time dependent. This leads to a very rich structure
of modes for the density contrast.
The perturbation equations read
\begin{eqnarray}
\ddot \delta &+& 2H \dot\delta = \frac{3}{2}\alpha H^2\left[ \delta\phi
+ \frac{1}{\alpha}\delta \right] - \delta E_{00}, \\
(\delta \phi)^{..} &+& 4 H (\delta \phi)^{.} + 2\alpha^2 H^2
(\delta\phi)= -2\alpha H^2 \delta,\\
(\delta E_{00})^. &+& 4H(\delta E_{00}) + 4(\delta H)E_{00} =
-3\alpha H^2 (\delta \phi)^. - 3\alpha H^2 \frac{\beta}{t}\delta
\nonumber \\ 
&+&\frac{3\beta}{8}\left[ -t^{-2}(\delta\phi)^. +t^{-1}(\delta
\phi)^{..}\right] + 3H\frac{\beta}{t}(\delta \phi)^. -
\frac{1}{2}\beta^2 t^{-2}\dot\delta. 
\end{eqnarray}
This system of equations possesses  power law solutions, which we
derive up to order $\alpha^2$. Defining each mode by  
the ansatz
\begin{equation}
\delta = \delta_{i} t^{a_i}, \delta\phi = \delta\phi_{i} t^{b_i}, 
\delta E_{00}= \delta E_{00}^i t^{c_i}
\end{equation}
we find that
\begin{equation}
a_i = b_i = c_i + 2
\end{equation}
for all modes. 
There are two types of modes. Let us first discuss the complete
solutions
of the coupled differential equations.
We find that there are three modes corresponding to
the exponents
\begin{equation}
a_{2/3}=\frac{2}{3}-\frac{11624}{7875}\alpha^2,\
a_{-1}=-1-\frac{1036}{375}\alpha^2,\
a_{-2/3}=-\frac{2}{3}+\frac{808}{225}\alpha^2
\end{equation}
The  modes  $\delta_{2/3}$ and $\delta_{-1}$ are deformations of the
FRW modes due to the scalar field. In particular we find that
$a_{2/3}\approx 0.55$. 
Thus, the growth of fluctuations is smaller, than in the normal matter
dominated epoch.  The perturbation of the
bulk scalar field is growing with the same exponent as the density
constrast and $\delta E_{00}$ is rapidly decaying with an 
exponent $c = a-2 \approx -1.45$. 

In addition to the growing mode, there are two decaying modes. One of
these decaying modes is a modification of the usual $a=-1$ decaying
mode. We obtain $a = -1.23$. Thus, the mode is decaying even faster 
than normal. The other decaying mode  corresponds to  $a=-0.37$.

On top of these three modes there are seven  homogeneous modes
corresponding to zero modes of the various differential operators
appearing
in the three coupled equations. Let us discuss them in some detail
as they have remarkable properties.
The homogeneous modes of the perturbed Raychaudhuri equation have
characteristic
exponents
\begin{equation}
\tilde a_{2/3}= \frac{2}{3}+\frac{16}{225}\alpha^2,\ \tilde
a_{-1}=-1+\frac{64}{225}\alpha^2
\end{equation}
Notice that these modes are two deformations of the usual FRW modes.
The scalar field potential lifts the degeneracy between the doublets of
$a=2/3$ and $a=-1$ modes. We find that $\tilde a_{2/3}\approx 0.67$
very close to the FRW value. That the growing mode is larger than 
the one found in general relativity comes from the fact that the
matter dominated universe is expanding {\it slower} in the brane world
model. Therefore, overdensities, which want to contract but have 
to compete with the cosmological expansion, can grow more easily 
in the brane world model. 

The homogeneous modes of the Klein-Gordon equation are characterized
by the exponents
\begin{equation}
a_0=-\frac{8}{15}\alpha^2,\ a_{-5/3}=-\frac{5}{3}+\frac{56}{45}\alpha^2,
\end{equation}
corresponding to an almost constant mode and a mode which decreased very
fast.

The $E_{00}$ equations comprises three differential operators.
The first order differential operator acting on $\delta E_{00}$ leads
to
a single mode with
\begin{equation}
a_{-2/3}=-\frac{2}{3}+\frac{32}{45}\alpha^2.
\end{equation}
The differential operator acting on $\delta\phi$ leads to two modes
\begin{equation}
\tilde a_0=0,\ a_{-10}= -10+\frac{512}{135}\alpha^2.
\end{equation}
Eventually the operator acting on $\delta$ leads to
\begin{equation}
\bar a_0=\frac{10}{27}.
\end{equation} 
Notice that in the epoch of accelerated expansion the structures get
frozen in. We have thus described the perturbation modes in the four
cosmological eras. 

In conclusion, we have found two the growing modes, with exponent
$\tilde a_{2/3}\approx 0.67$ and $a_{2/3} \approx 0.55$. 
The 00--component of the Weyl--tensor is always decaying, whereas perturbations in $\phi$ 
follow the density contrast. It is instructive to compare these 
results to the ones found in Brans--Dicke theory. Here the exponent 
of the growing and decaying modes are modified as follows \cite{lobo}:
\begin{equation}
a_{+} = \frac{2}{3} + \frac{2}{3\omega} 
\hspace{0.5cm} {\rm and} \hspace{0.5cm} a_{-} = -1 - \frac{1}{3 \omega},
\end{equation}
where $\omega$ is the Brans--Dicke parameter. Note, that there are
more modes in these theories as well. Here we find  that the
growing and decaying modes are shifted due to the presence of a scalar
field which couples to gravity in a way which differs from Brans-Dicke
theory.

\subsubsection{Observational consequences}
Although we have focused on the large wavelength limit only, we
can draw some (qualitative) conclusions from our findings. First of
all, which of the two growing modes in the matter dominated epoch 
appear, depends on the initial conditions imposed in the early
universe. It is likely that both modes are generated. If the 
mode with exponent $a \approx 0.67$ is not generated and 
the power spectrum is normalized to the observations today (such as
to galaxy clusters), the brane world model would have much more power in
the matter perturbations than one would expect in normal quintessence
models, for example. Thus, in this case there should be more galaxy
clusters at high redshift in the brane world model than in normal 
Einstein gravity with comparable matter density (with a cosmological 
constant or quintessence field). If the mode with exponent 
$a \approx 0.67$ contributes significantly to perturbations, 
then the power in the perturbations in both theories would be
similar. However, much more work is needed in order to make 
this more concrete. 

Another consequence is a modification of the spectrum of anisotropies 
in the cosmic microwave background radiation (CMB). This is because 
the distance to the last scattering surface (LSS) will be modified 
due to the slower expansion in the matter dominated epoch. As a
result, the first peak will be shifted to larger angular scales. 
However, we believe this effect to be small to be detected.
In addition to this effect, it is 
conceivable that isocurvature modes (between the Weyl fluid and
radiation, for example) might survive quite long in the radiation era, 
leaving their imprint in the CMB. To make more concrete predictions,
it is necessary to go beyond the small $k$--limit, which we considered
in this paper. This involves a study of the bulk gravitational field,
as well as perturbations in the bulk scalar field away from the
brane \cite{braneworldperturbations}. The calculations are very difficult 
and a detailed discussion of the bulk equations is beyond the scope of 
this paper and will be presented elsewhere.

\section{Conclusions}
We have presented a discussion of cosmological perturbations in brane
world scenarios with a bulk scalar field. For a model motivated from 
SUGRA in singular spaces, we were able to find solutions of the
perturbation equations in the large scale limit. As is the case
with brane world scenarios of the Randall--Sundrum type, the
pertrubation equations on the brane are not closed. Instead one has
also to solve the bulk equations, which is a very difficult task in
general. 

Nevertheless, our findings indicate that there are considerable
differences to the Randall--Sundrum scenario as well as to usual
scalar--tensor theories in four dimensions. The growth of structures
is different than in these models. A more elaborate investigation 
of perturbations, in particular of the bulk perturbations, would be necessary
in order to make definitive predictions concerning cluster abundance, 
large scale structure and the anisotropies in the CMB.

There are other open questions, which our work leaves. As discussed in
\cite{BDIII}, in models based on SUGRA in singular spaces we
expect a second brane in the bulk at some distance from our brane
universe. We have not addressed the influence of this mirror brane on the 
structures on our brane universe. Similarly, we haven't discussed the 
dynamics of the radion in these models. The effects of these should
be encompassed in the Weyl tensor. We will, however, turn to these
questions in future work. 

\acknowledgments We are grateful to Brandon Carter, Jai-Chan Hwang, 
Zygmunt Lalak, Roy Maartens, Andrew Mennim and David Wands for 
discussions. This work was supported in part by the Deutsche 
Forschungsgemeinschaft (DFG) and PPARC (UK).

\begin{appendix}
\section{Supersymmetric Backgrounds}
\subsection{BPS configurations}
In this appendix we will consider configurations which preserve
supersymmetry in the bulk. To do so we will look for Killing
spinors satisfying the identities
\begin{equation}
\delta_{\epsilon}\psi_{\mu i}=0,\ \delta_{\epsilon}\lambda_i^x=0
\end{equation}
where $\psi_{\mu i}$ is the gravitino spinor $(i=1,2)$ and
$\lambda_i^x$ belongs to the vector multiplet comprising the
scalar field $\phi^x$.
This leads to the first order equations
\begin{equation}
D_{\mu} \epsilon_i+\frac{i}{8}\gamma_{\mu}W Q_{ij}\epsilon^j=0
\end{equation}
and
\begin{equation}
i\gamma^{\mu}\partial_{\mu}\phi^x\epsilon_i + W^{,x}Q_{ij}\epsilon^j=0
\label{lam}
\end{equation}
where $Q_{i}^j= Q_a(\sigma^a)_{i}^j$ and  $Q_aQ^a=1$.
Define the vector $p_a$ such that
\begin{equation}
\partial_a \phi^x =p_a \partial_z\phi^x
\end{equation}
and the matrix
\begin{equation}
\Gamma=\frac{p_a\gamma^a}{\sqrt {p^2}}
\end{equation}
such that $\Gamma^2=1$.
Rotation invariance implies that the only non-zero components are $p_z=1$ and $p_t$.
The unknown component $p_t$ is fixed by the boundary condition.
Now the spinors $\epsilon_i$ can be split into positive and
negative chiralities
\begin{equation}
\epsilon_i=\epsilon_i^++\epsilon_i^-
\end{equation}
satisfying
\begin{equation}
\epsilon_i^{\pm}=\pm i \Gamma Q_{ij}\epsilon^{\pm j}
\label{pro}
\end{equation}
At each point this select the  chirality defined by $\Gamma$, i.e.
we preserve one half of the supersymmetries.
Now (\ref{lam}) is satisfied provided
\begin{equation}
\partial_z\phi^x=\pm \frac{1}{\sqrt {p^2}}W^{,x}
\label{BP}
\end{equation}
This is the BPS equation for supersymmetric configurations.
Now choosing the positive
sign, we find that the boundary condition at $z=0$ is
automatically satisfied provided
\begin{equation}
T\sqrt{{p^2}}=1
\end{equation}
This leads to the vanishing
\begin{equation}
\Delta\Phi_2=0
\end{equation}

Let us come to the bulk evolution of the Killing spinors. They satisfy
\begin{equation}
D_\mu \epsilon_i+\frac{i}{8}\gamma_{\mu}W Q_{ij}\epsilon^j=0
\label{e}
\end{equation}
This is a first order differential equation which can be solved
provided the integrability condition
\begin{equation}
[D_a,D_b]\epsilon_i^+=R_{abcd}\frac{\gamma^{cd}}{8}\epsilon_i^+
\end{equation}
is fulfilled.
Now one gets
\begin{equation}
[D_a,D_b]\epsilon_i^+=-\frac{1}{64}W^2 \gamma_{ab}+\frac{1}{8}\frac{p_a\gamma_b-p_b\gamma_a}{\sqrt{p^2}}
W^{,x}W_{,x}\Gamma)\epsilon_i^+
\end{equation}
from which we deduce that
\begin{equation}
R_{abcd}=-\frac{1}{16}W^2 (g_{ac}g_{bd}-g_{ad}g_{bc})+\frac{1}{4p^2}W^{,x}W_{,x}(
p_ag_{b[d}p_{c]}-p_bg_{a[d}p_{c]})
\label{R}
\end{equation}
For $W_{,x}=0$, i.e. at the critical points of the superpotential
one recognizes the Riemann tensor of $AdS_5$. More generally we
find that the background geometry which preserves supersymmetry
in the bulk is such that the Weyl tensor vanishes
\begin{equation}
W_{abcd}=0
\end{equation}
as the Weyl tensors form, in the set of curvature tensors, the complement
to the antisymmetrized product of the metric tensor with itself
or  with a symmetric tensor.
This implies that the bulk geometry is conformally flat and can
be written as
\begin{equation}
ds^2=e^{2A(z,t)}(-dt^2 +dz^2 +dx_idx^i)
\end{equation}
This leads to the vanishing of the projected Weyl tensor
\begin{equation}
E_{ab}=0
\end{equation}
in the bulk.
Moreover the bulk and brane expansion rates coincide
\begin{equation}
H_B=H.
\end{equation}
Notice that (\ref{R})
leads to the Ricci tensor
\begin{equation}
R_{ab}=-\frac{W^2}{4}g_{ab}+\frac{W^{,x}W_{,x}}{4p^2}(p^2g_{ab}+3p_ap_b)
\end{equation}
and the curvature scalar
\begin{equation}
a^2 R=-\frac{5W^2}{4}+2W^{'x}W_{,x}
\end{equation}
This implies that Einstein equations
\begin{equation}
R_{ab}-\frac{R}{2}g_{ab}=T_{ab}
\end{equation}
are satisfied with
\begin{equation}
\partial_z A=-\frac{U_B}{4}
\label{A}
\end{equation}
Notice that  (\ref{A}) and (\ref{BP}) are the BPS equations which lead
to the accelerating universe.
Solutions of these equations satisfy the Einstein equations and the
Klein-Gordon equation.

Let us now consider the second brane where supersymmetry is not
broken.
Now the Killing spinors for the supersymmetric $T=1$ case
satisfy the projection equation
\begin{equation}
\epsilon_i^{\pm}=\pm i \gamma^5 Q_{ij}\epsilon^{\pm j}
\end{equation}
as $p_t=0$. This is not compatible with (\ref{pro}) as $\Gamma\ne
\gamma^5$.
The only solution is therefore $\epsilon=0$, i.e. supersymmetry is
completely broken by the non-supersymmetric brane with $T\ne 1$.
Notice that the breaking of supersymmetry is global due to the presence of
two boundaries respecting incompatible supersymmetries.

So we have shown that breaking supersymmetry on the brane leads to
broken $N=0$ background configurations which still satisfy a system of
two first order BPS conditions. 

\subsection{Breaking Conformal Flatness}

Let us now consider 
the case where matter and radiation are present on the brane. 
We will show that  one cannot 
deform the bulk geometry in such a way that the boundary conditions
are satisfied. This implies that matter on the brane breaks the
conformal flatness of the bulk.

Let us
perform a small change of coordinates in the bulk
\begin{equation}
\tilde x=x-\xi
\end{equation}
inducing  a variation of the bulk metric  is
\begin{equation}
\delta g_{ab}=\partial_a\xi_b+\partial_b\xi_a
\end{equation}
and of the scalar field
\begin{equation}
\delta(\partial_a\phi)=\partial_c\phi\partial_a\xi^c, \
\delta(\partial_a\partial_b\phi)=\partial_c\partial_a\phi \partial_b\xi^c
+\partial_c\partial_b \phi \partial_a\xi^c
\end{equation}
We also need to evaluate the variation
\begin{equation}
\delta (\partial_z g_{ab})=\partial_z\xi^{c}\partial_cg_{ab}
+\partial_a\xi^c\partial_z g_{cb}+ \partial_b\xi^c\partial_zg_{ca}
\end{equation}
The boundary equation $\delta(\partial_z\phi)\vert_0=0$ leads to
\begin{equation}
\partial_z\xi^z\vert_0 +p_t \partial_z\xi^t=0
\end{equation}
We then find that 
\begin{equation}
\delta(\partial_zg_{aa})\vert_0=2\partial_a\xi^a\partial_z g_{aa}\vert_0
\end{equation}
with no summation involved. 
The metric boundary condition at the origin are 
modified according to
\begin{equation}
\delta(\frac{\partial_z g_{aa}}{g_{aa}})\vert_0=\frac{\delta(\partial_z g_{aa})}{g_{aa}}\vert_0-
\frac{\partial_z g_{aa}}{g_{aa}}\vert_0\frac{\delta g_{aa}}{g_{aa}}\vert_0
\end{equation}
implying that
\begin{equation}
\delta(\frac{\partial_z g_{aa}}{g_{aa}})\vert_0=
-\frac{U_B}{2}\partial_a g^{aa}\xi_a\vert_0
\end{equation}
with no summation involved.
The $g_{ii}$ boundary condition then reads
\begin{equation}
\frac{U_B}{2}\partial_i g^{ii}\xi_i\vert_0=\frac{\rho}{6}
\end{equation}
As the background is $x^i$ independent this cannot be satisfied unless
$\rho=0$.
This proves that the bulk metric cannot be smootly obtained from the
matterless case by performing a change of coordinates in the bulk.
In particular conformal flatness is broken leading to an explicit
breaking of supergravity by matter on the brane.

\end{appendix}


\begin{thebibliography}{99}
\bibitem{Polchinski} J. Polchinski, {\it String Theory}, 
Vol.1 + Vol.2, Cambridge University Press (1999)
\bibitem{HW1} E. Witten, Nucl. Phys. B {\bf 471}, 135 (1996)
\bibitem{HW2} A. Lukas, B. Ovrut, K. Stelle, D. Waldram, 
Phys. Rev. D {\bf 59}, 086001 (1999)
\bibitem{RSII} L. Randall, R. Sundrum, 
Phys. Rev. Lett. {\bf 83}, 4690 (1999)  
\bibitem{RScosmo} 
J.M. Cline, C. Grojean, G. Servant, Phys. Rev. Lett {\bf 83}, 4245 (1999);
C. Csaki, M. Graesser, C. Kolda, J. Terning, 
Phys. Lett. B {\bf 462},34 (1999);
T. Shiromizu, K. Maeda, M. Sasaki, Phys. Rev. D {\bf 62}, 024012 (2000); 
P. Binetruy, C. Deffayet, U. Ellwanger, D. Langlois, Phys. Lett. B. 
{\bf 477}, 285 (2000); P. Kraus, JHEP 9912, 011 (1999);
L. Anchordoqui, C. Nunez, K. Olsen, JHEP 0010, 050 (2000);
E. Flanagan, S. Tye, I. Wasserman, Phys.Rev.D {\bf 62}, 044039 (2000); 
R. Maartens, D. Wands, B. Bassett, I. Heard, Phys.Rev.D {\bf 62}, 
041301 (2000); E. Copeland, A. Liddle, J. Lidsey,
Phys.Rev.D {\bf 64}, 023509 (2001); G. Huey, J. Lidsey, 
Phys.Lett.B {\bf 514}, 217 (2001); 
A. Davis, C. Rhodes, I. Vernon, hep-ph/0107250;
S. Davis, W. Perkins, A. Davis, I. Vernon, Phys.Rev.D {\bf 63},083518 (2001); 
R. Maartens, V. Sahni, T.D. Saini, Phys.Rev.D {\bf 63}, 063509 (2001); 
M. Santos, F. Vernizzi, P. Ferreira, hep-ph/0103112; 
A. Campos, C. Sopuerta, Phys.Rev.D {\bf 63}, 104012 (2001) and hep-th/0105100;
V. Sahni, M. Sami, T. Souradeep, gr-qc/0105121;
S. Mizuno, K. Maeda, hep-ph/0108012
\bibitem{robert} V. Mukhanov, H. Feldman, R. Brandenberger, 
Phys.Rept. {\bf 215}, 203 (1992)
\bibitem{braneworldperturbations}
H. Kodama, A. Ishibashi, O. Seto, Phys.Rev.D {\bf 62}, 064022 (2000); 
R. Maartens, Phys.Rev.D {\bf 62}, 084023 (2000);
S. Mukohyama, Phys. Rev. D {\bf 62}, 084015 (2000), 
Class.Quant.Grav. {\bf 17}, 4777 (2000) and hep-th/0104185;
C. van de Bruck, M. Dorca, R. Brandenberger, A. Lukas, 
Phys.Rev.D {\bf 62}, 123515 (2000);
D. Langlois, Phys.Rev.D {\bf 62}, 126012 (2000) 
and Phys.Rev.Lett. {\bf 86}, 2212 (2001);
C. Gordon, R. Maartens, Phys.Rev.D {\bf 63}, 044022 (2001);
N. Deruelle, T. Dolezel, J. Katz, Phys.Rev.D {\bf 63}, 083513 (2001);
N. Deruelle, J. Katz, gr-qc/0104007;
S.W. Hawking, T. Hertog, H.S. Reall, Phys.Rev.D {\bf 63}, 083504 (2001); 
A. Neronov, I. Sachs, Phys. Lett. B. {\bf 513}, 173 (2001);
D. Langlois, R. Maartens, M. Sasaki, D. Wands, 
Phys.Rev.D {\bf 63}, 084009 (2001);
M. Dorca, C. van de Bruck, Nucl. Phys. B {\bf 605}, 215 (2001);
C. van de Bruck, M. Dorca, hep-th/0012073;
H. Bridgman, K. Malik, D. Wands, astro-ph/0107245;
K. Koyama, J. Soda, Phys.Rev.D{\bf 62}, 123502 (2000) and hep-th/0108003
\bibitem{singular} E. Bergshoeff, R. Kallosh, A. Van Proeyen, 
JHEP {\bf 0010}, 033, (2000),  A. Falkowski, Z. Lalak and S. Pokorski
{\it Phys. Lett.} {\bf B491} (2000) 172,  
R. Altendorfer, J. Bagger and D. Nemeschansky, Phys.Rev.D {\bf 63},
125025 (2001); 
T. Gherghetta, A. Pomarol, Nucl.Phys.B {\bf 586}, 141 (2000).
\bibitem{BDI} P. Brax, A.C. Davis, Phys.Lett.B {\bf 497}, 289 (2001) 
\bibitem{BDII} P. Brax, A.C. Davis, JHEP 0105, 007 (2001)
\bibitem{bulkscalar} D. Langlois, M. Rodriquez-Martinez, hep-th/0106245;
S.C. Davis, hep-th/0106271; 
C. Charmousis, hep-th/0107126
\bibitem{self-tune} N. Arkani-Hamed, S. Dimopoulos, N. Kaloper 
and R. Sundrum, Phys. Lett. B. {\bf 480}, 193 (2000);
S. Kachru, M. Schulz and E. Silverstein, Phys. Rev. D. {\bf 62},
045021 (2000).
\bibitem{mennim} A. Mennim, R. Battye, Class.Quant.Grav. {\bf 18}, 
2171 (2001); K. Maeda, D. Wands, Phys.Rev.D {\bf 62}, 124009 (2000) 
\bibitem{vdB} C. van de Bruck, M. Dorca, C.J.A.P. Martins, M. Parry, 
Phys.Lett.B {\bf 495}, 183 (2000) 
\bibitem{Nordtvedt}
T. Damour, K. Nordtvedt, Phys.Rev.Lett. {\bf 70}, 2217 (1993), 
and Phys.Rev.D {\bf 48}, 3436 (1993)
\bibitem{Barrow} J. Barrow, Phys.Rev.D {\bf 47}, 5329 (1993)
\bibitem{wetterich}
C. Wetterich, Nucl. Phys. B {\bf 302}, 668 (1988)
\bibitem{Hawking} S.W. Hawking, Astrophys. Journ. {\bf 145}, 544 (1966)
\bibitem{Lyth} D. Lyth, E. Stewart, Astrophys. Journ. {\bf 361}, 343
(1990); J.-C. Hwang and E. Vishniac, ApJ {\bf 353}, 1 (1990); 
J.-C. Hwang, ApJ {\bf 380}, 307 (1990);
\bibitem{Liddle} A. Liddle, D. Lyth, {\it Cosmological Inflation and 
Large Scale Structure}, Cambridge University Press (2000)
\bibitem{lobo} E. Gaztanaga, J.A. Lobo, Astrophys. Journ. {\bf 548}, 
47 (2001)
\bibitem{BDIII} P. Brax, A. Davis, Phys. Lett. B {\bf 513}, 156 (2001)
\end{thebibliography}
\end{document}